\documentclass[a4paper]{article}

\usepackage{INTERSPEECH2021}
\usepackage{CJK}

\title{Microphone Array Generalization for Multichannel \\ Narrowband Deep Speech Enhancement}
\name{Siyuan Zhang, Xiaofei Li}
\address{
  School of engineering, Westlake University, Hangzhou, China \\ 
Institute of Advanced Technology, Westlake Institute for Advanced Study, Hangzhou, China}
\email{zhangsiyuan@westlake.edu.cn, lixiaofei@westlake.edu.cn}

\begin{document}

\maketitle
\begin{abstract}
  This paper addresses the problem of microphone array generalization for deep-learning-based end-to-end multichannel speech enhancement. We aim to train a unique deep neural network (DNN) potentially performing well on unseen microphone arrays. The microphone array geometry shapes the network's parameters when training on a fixed microphone array, and thus restricts the generalization of the trained network to another microphone array. To resolve this problem, a single network is trained using data recorded by various microphone arrays of different geometries. We design three variants of our recently proposed narrowband network to cope with the agnostic number of microphones. Overall, the goal is to make the network learn the universal information for speech enhancement that is available for any array geometry, rather than learn the one-array-dedicated characteristics. The experiments on both simulated and real room impulse responses (RIR) demonstrate the excellent across-array generalization capability of the proposed networks, in the sense that their performance measures are very close to, or even exceed the network trained with test arrays. Moreover, they notably outperform various beamforming methods and other advanced deep-learning-based methods. 
\end{abstract}

\section{Introduction}

Microphone arrays are extensively applied in intelligent speech communication systems. Multichannel speech enhancement, owing to its additional spatial information, is widely studied in recent years and tends to have superior performance relative to monaural speech enhancement. Beamforming \cite{beamforming} is a classic multichannel speech enhancement model preserving the desired signal from the target direction and suppressing signals from other directions. Different in the manners of utilizing spatial information, recent studies on multichannel speech enhancement DNNs can be divided into two categories: \romannumeral1) monaural deep masking or mapping \cite{overview} followed by beamforming \cite{NNGEV},\cite{Masking-MVDR}; \romannumeral2) directly learning spatial information by inputting multichannel signals or cross-channel features into the network, and performing end-to-end speech enhancement. Methods in the first class have become the most popular front-ends for automatic speech recognition (ASR) \cite{E2EASR}. In addition, they are free from the microphone array generalization problem, since both monaural speech enhancement DNN and beamforming can be flexibly applied on any array configurations. Especially, \cite{CSM-ASR} and \cite{SM-Deverberation} further strengthen the array generalization capability by exhaustively employing each microphone in an array as a reference microphone. However, these approaches solely focus on learning monaural speech T-F information channel by channel with DNNs and the performance strongly depends on the estimation accuracy of beamforming parameters. We thus believe that simultaneously exploiting the spatial and monaural information with DNN could further improve the estimation of speech on target direction, as is done in the end-to-end time-domain methods \cite{timeDAE,timeFCN} and T-F domain methods \cite{CRNN,CONSISTENCY}. Neural beamforming takes the multichannel signals as input and explicitly performs as a beamformer \cite{DBN,CA-UNET,DBN-ASR} and sometimes it is trained to leverage the spatial cues, such as interaural phase/level difference in \cite{BSS-REV,All-neural,DCRNN}. These methods apparently have the microphone array generalization problem which, however, have not been considered in their original papers. Very recently, the FaSNet \cite{fasnet} integrated with a transform-average-concatenate (TAC) \cite{permutation} was proposed to process signals with varying number of microphones, where the TAC module together with cross-correlation functions between channels were employed to learn the spatial information. In \cite{distributed}, a self-attention network is used to exploit the spatial information with a varying number of microphones.  Both these two methods, i.e. \cite{permutation} and \cite{distributed}, alternate the temporal processing layers and spatial processing layers, which is arguably problematic for learning some temporal-spatial-combined information, such as the directional continuity of speakers and the spatial diffuseness of noise.  

\begin{figure*}[htbp]
\begin{center}
\includegraphics[scale=0.80]{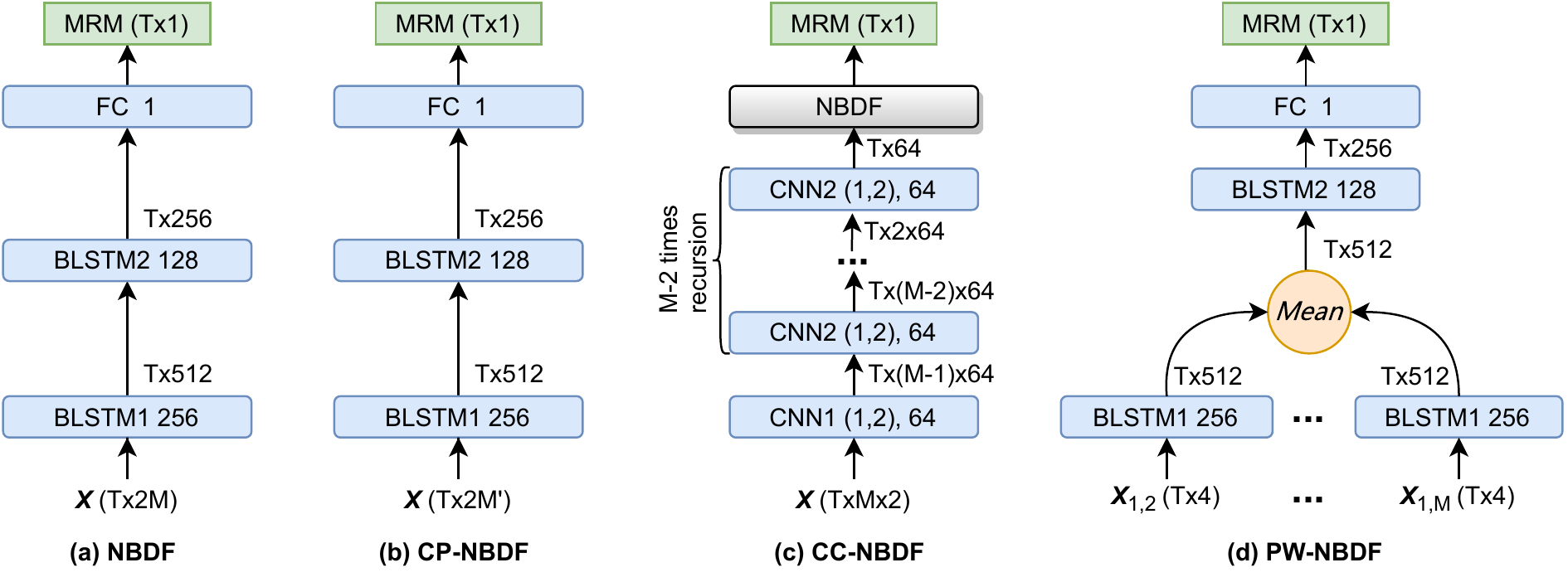}
\end{center}
\caption{Diagram of the basic NBDF and proposed three variants of NBDF network architectures.}
\label{networks}
\end{figure*}

In this work, we propose a new speech enhancement network being able to work with any microphone arrays. The importance of microphone array generalization is two-fold. First, the spatial correlations of multichannel signals are array-dependent. Previously, the training and test data are always recorded with the same microphone array. However, recording a large dataset with a new microphone array is time and budget-consuming. Second, DNN models with great microphone array generalization capability are favorable in Cloud applications and robot group intelligence, where recording end-devices are different. Our proposed networks, once trained, are able to enhance multichannel signals without requiring any prior knowledge of microphone arrays. 

In our previous works \cite{subband,NBDF}, a narrow band deep filtering (NBDF) method was proposed, which inputs the frequency-wise multichannel signals to a long short-term memory (LSTM) network, and predicts the clean speech at the corresponding frequency. NBDF has achieved excellent speech enhancement performance in terms of both human perception and ASR, thus it is used as the baseline network in this paper. We propose three variants of NBDF in this work to account for the varying number of microphones, namely channel-padding NBDF (CP-NBDF), channel-convolutional NBDF (CC-NBDF), and pair-wise NBDF (PW-NBDF). A large number of virtual microphone arrays with simulated RIRs are used for training. Multichannel signals are generated by convolving RIRs with single-channel speech clips. By using the toolkit presented in \cite{noise-generator} and \cite{wind-generator}, multichannel spatially diffuse babble, white and spatially correlated wind noise are generated respecting geometries of those arrays. The multichannel speech enhancement network normally recovers the monaural clean speech of a reference channel. We propose a simple and effective network training scheme to allow the users to specify a reference channel for their own microphone array during test. Besides, a magnitude augmentation scheme is proposed to account for the fact that different microphone arrays could have different configurations in terms of recording volume or frequency response. Overall, combining all of these, the network is trained to learn the universal information for speech enhancement that is available in any microphone array. Experiments demonstrate that our proposed networks have excellent generalization capability to various unseen simulated and real arrays. Moreover, they achieve better speech enhancement performance than various beamforming methods and the advanced FaSNet-TAC method \cite{permutation}.
\vspace{-0.1cm}
\section{System description}
\subsection{Signal model and narrow band deep filtering recap}
The multichannel microphone signal $ X $ is the mixture of multichannel clean speech $ S $ and noise $ N $ represented in the short-time Fourier transform (STFT) domain. 
\vspace{-0.1cm}
\begin{equation}
X_m(k,t)=S_m(k,t)+N_m(k,t). \label{1}
\end{equation} 
where $m \in [1,M], k\in[1,K], t\in[1,T]$ are the microphone, frequency and time indices, respectively. $T$ and $K$ are the total number of frames and frequency-bins. $M$ denotes the number of microphones. This work focuses on studying speech enhancement on different microphone arrays to predict the speech $S_r(k,t)$ of one selected reference channel, thence the reverberation is not suppressed even if presenting in $\hat{S}_r(k,t)$.

NBDF (Fig.1 (a)) is a DNN processing multichannel speech frequency-wisely. Uncoupling the inter-frequency dependency from full-band signal makes the network focus on learning the narrow-band spatial and temporal information to recognize speech components on each T-F unit. Besides, by solely exploiting narrowband information, the NBDF network generalizes well to new speakers and new noise scenarios with unseen spectral patterns, which naturally matches with our objective, such as the Cloud applications. NBDF network is composed of two stacked Bi-directional LSTM (BLSTM) layers and one fully connected (FC) layer with sigmoid activation before the output, shown in Fig. 1 (a). Different frequencies of one signal are independently processed by the network. The frequency-wise input $ \mathbf{X} \in \mathbb{R}^{T \times 2M} $, is the concatenation of multichannel real and imaginary parts of STFT coefficients channel by channel. The target, i.e. monaural magnitude ratio mask (MRM) sequence $ \in \mathbb{R}^{T \times 1} $, is the ratio of magnitudes between clean speech and noisy mixture at the reference channel. 
During inference, the predicted MRM is multiplied with the noisy mixture of the reference channel to restore the magnitude of clean speech. The phase of the recovered signal remains the same as the noisy mixture. Each input sample is normalized by dividing the mean of the STFT magnitude of all frames at the reference channel. We take the mean square error of MRM as the training loss. The enhanced reverberant reference channel signal $\hat{s_r}(t)$ is finally obtained with inverse STFT.

\vspace{-0.1cm}
\subsection{Inner microphone array generalization}
\vspace{-0.1cm}
To achieve the cross-array generalization, it's necessary to first resolve the inner-array generalization, namely the generalization problem for one constant microphone array. This includes two aspects: i) the reference channel selection mechanism. During test, different reference channels can be used with the already-trained network. The optimal reference channel can be selected by the human users based on their prior knowledge or by an automatic selection algorithm \cite{improvedMVDR}. The network then predicts the clean speech of the selected reference channel; ii) the network's immunity to the microphone permutation. The rationale is that the spatial information should be conveyed by the correlation between microphones rather than by a certain recording permutation of the microphone array.
These two aspects are also very important when the network is tested on an unseen microphone array, as its reference channel and permutation are completely irrelevant to the training arrays. A network trained with a constant reference channel and a fixed permutation, will overfit to such setup and hardly generalizes to others.

We propose a novel training scheme to account for the addressed aspects. Regarding reference channel, a trivial way is to encode the reference channel selection as a one-hot vector and input it to the network along with the signals.  This one-hot vector will specify the reference channel index during test. We simplify this scheme by using a constant one-hot vector that always specifies the first channel as the reference channel and shift it to the first channel, namely the first two elements in $\mathbf{X}$. The constant one-hot vector can be omitted in this way. Rest channels are shuffled (independently for each sample) to make the network trained with various permutations. As a consequence, the network is able to recognize the reference channel (always the first input channel) in the basic NBDF network, CC-NBDF, and CP-NBDF and to be invariant to the permutation.
\vspace{-0.15cm}
\subsection{Three variants of NBDF}
\vspace{-0.2cm}
The most straightforward way to make the NBDF network adapt to microphone arrays with varying number of microphones is to set an upper bound for the number of microphones, say $M'$ (preset eight in this paper) in channel-padding NBDF (shown in Fig. 1 (b)). The network is trained on various arrays with maximum $M'$ microphones. When $M<M'$, the $M$ channels are first organized using the training scheme in Section 2.2 and the last $M'-M$ channels are padded with zeros. It always learns the global spatial information as in basic NBDF. This mechanism partially solves the varying number of microphones problem, since it can not process more than $M'$ channels.

In channel-convolutional NBDF, as is done in \cite{CRNN}, the inter-channel information can be locally learned by time-distributed convolutional neural networks (CNN)s, and correspondingly the channel-dimension can be ablated with multilayer CNNs. We proposed a channel convolutional NBDF network, as shown in Fig. 1 (c). The input, $ \mathbf{X} \in \mathbb{R}^{T \times M \times 2} $, is convolved by the first CNN layer to fixed $N$ feature maps and then is transferred to $M-2$ consecutive shared CNN layers. Channel convolutional module is followed by a basic NBDF model. For the dual-channel case, only the first CNN layer functions. All channels of input are organized with the above training scheme.  Comparing to the basic NBDF network which learns global spatial information, the CC-NBDF uses CNNs to recursively learn the accumulated local inter-channel information, which may lack some global channel dependencies.

In pair-wise NBDF shown in Fig. 1 (d), each microphone pair is composed of the reference microphone and one of the other microphones: $ \mathbf{X}_{1,m}\in\mathbb{R}^{T\times4}, \ m\in[2, M]$, where $1$ in subscript means the first channel is reference channel. The input sequence $ \mathbf{X} \in \mathbb{R}^{T \times 2M} $ is uncoupled as $M-1$ dual-channel narrowband sequences which are parallelly input to the shared LSTM, i.e. BLSTM 1 in Fig. 1 (d). This works for any array with a different number of microphones and free from the permutation problem owing to pair-wise mapping. The $M-1$ outputs of this LSTM layer are averaged and then sent to BLSTM 2 and FC layer same as the basic NBDF network. Similar to channel convolutional NBDF, it learns inter-channel correlation pair-wisely but differently learns the global channel relation from the mean-pooled feature. The good point is that all microphone pairs are formed with the reference channel, which provides sufficient information for mask estimation at the reference channel. 

\vspace{-0.2cm}
\subsection{Magnitude augmentation}
\vspace{-0.1cm}
Each microphone has its own characteristics such as its unique frequency response, recording volume, and noise floor. It is reflected by the different magnitude levels while recording the same signal. To take this effect into account, during training, we multiply STFT magnitude for each microphone at each frequency with a randomly chosen factor. This data augmentation technique largely increases the data diversity in terms of the relative magnitudes between microphones and may improve the generalization capability while testing unseen array. 

\begin{table}[t]
\centering
\caption{Speech enhancement results on a constant array (SIMU 6-CH LA). SNR is in the range of [-5,0] dB.}
\tabcolsep0.025in
\begin{tabular}{l| c c c}

Model                         & PESQ   & STOI (\%)  & SDR (dB) \\ 
\hline
Unprocessed                   & 1.68   & 63.9   & -2.4    \\
Constant-reference-permutation & 2.67   & 78.2  &  10.6    \\
Constant-permutation          & 2.64   & 77.3     &  10.5 \\
Proposed training scheme      & 2.62   & 76.5     &  10.3 \\
\end{tabular}
\label{tab: proposed training scheme group}
\vspace{-0.5cm}
\end{table}

\begin{table*}[t]
\scriptsize
\centering
\caption{Speech enhancement results of microphone array generalization. SNR is uniformly distributed in the range of [-5,0] dB.}
\label{tab:Result for different microphone array generalization}
\tabcolsep0.04in
\begin{tabular}{c l c |c c c |c  c c|c  c c|c c c |c c c c | c c }   
\toprule[1.25pt]
& PESQ / STOI (\%) / SDR (dB)  & Model size  & \multicolumn{3}{c|}{2-CH }  &  \multicolumn{3}{c|}{4-CH CA }  &  \multicolumn{3}{c|}{4-CH LA } &  \multicolumn{3}{c|}{6-CH CA } &  \multicolumn{3}{c}{6-CH LA } \\ 
   \cmidrule(lr){1-2}  \cmidrule(lr){3-3}  \cmidrule(lr){4-6}  \cmidrule(lr){7-9} \cmidrule(lr){10-12} \cmidrule(lr){13-15} \cmidrule(lr){16-18}
     & Unprocessed                  &   - &  1.70& 62.4 & -2.4& 1.70& 61.5&-2.5 & 1.69 & 63.1&-2.3&	1.70 &  61.6&-2.5&	1.70     &63.0 &-2.3\\ 
     & BeamformIt \cite{beamformit} & -     &  1.81 & 63.0& -1.1  & 1.72& 59.5 &-2.7 & 	1.87 & 63.2	& -0.5&1.78 & 60.9 &-1.6& 1.88  &62.8 &-0.3\\
     & Oracle MVDR                  & -     &  1.83	&	64.7&-0.5 & 1.84 & 64.9 &-0.5 & 1.93& 68.3 &1.1&	1.94 &  66.7& 0.5 &	1.97     &	69.1 & 1.5 \\
     & FaSNet-TAC \cite{permutation}         & 2.75M &  2.36 &	\textbf{73.5} & \textbf{10.0} & 2.43& 74.0 &10.1 &2.40 & 75.3& \textbf{10.5} &	2.47&  74.7 & \textbf{10.4} &	2.43     &	75.7& \textbf{10.6}\\
     & CP-NBDF (prop.)                     & 1.21M & 2.53 & 72.8&9.5& 2.72 & 75.2&9.9&2.62& 74.9 &10.1&	2.75 & 75.7 &10.0&	2.64    &	75.1& 10.1\\
SIMU & CC-NBDF (prop.)                     & 1.33M & 2.51 & 72.1 & 9.4& 2.70  & 74.6&9.8&2.60 & 74.4&10.1&	2.74 & 75.1&9.9&	2.64    &	74.8&10.1 \\
      & PW-NBDF (prop.)                     & 1.19M & 2.49 & 73.0 &9.2& 2.73 & \textbf{76.1}&10.0&2.61 &  \textbf{75.8} &10.0& 2.79 & 76.9 &10.2&	2.67     & \textbf{76.6}&10.2 \\ 
     & PW-NBDF with MA (prop.)             & 1.19M & \textbf{2.56} &73.1&9.8& \textbf{2.77} & \textbf{76.1}&\textbf{10.2}&\textbf{2.66}  &  \textbf{75.8}&10.4&	\textbf{2.82} &  \textbf{77.0} &10.3&	\textbf{2.71}    & \textbf{76.6}&10.5 \\ 
     \cmidrule[0.25pt](lr){2-3} \cmidrule[0.25pt](lr){4-6} \cmidrule[0.25pt](lr){7-9} \cmidrule[0.25pt](lr){10-12} \cmidrule[0.25pt](lr){13-15} \cmidrule[0.25pt](lr){16-18}
     & One-array-dedicated          & 1.20M & 2.58  &	73.6&10.0& 2.69 & 76.0&9.0&2.63 & 76.0&10.2&	2.77& 76.4  &10.1&	2.65    &	75.8&10.2\\  
     & Microphone-number-dedicated  & 1.20M & 2.52 &	72.8 & 9.7 & 2.75 & 76.0 &10.1& 2.62  & 75.1&10.3&	2.78 & 76.3 &10.1&	2.66    &	75.8&10.2\\ 
     \midrule[0.75pt]
     & Unprocessed                   & - &  1.67 &	63.4 & -2.9 & 1.70  & 64.3 &-2.4& 1.67  &68.4 &-2.3&	1.67 & 62.7&-3.1&	1.66  &	68.0&-2.4 \\ 
     & BeamformIt \cite{beamformit}  & - &  1.73  &	64.3&-1.4& 1.72 & 63.9 &-1.7& 1.80  & 70.5 &-0.7&	1.75 &  63.6 &-1.8&	1.80  &70.4 &-0.9\\
     & Oracle MVDR                   & - &  1.84 &	69.9 &0.4& 1.88 &  72.7 &0.5& 2.09 &  \textbf{84.2} &3.0&	2.02  &77.4 &1.8&	2.16     &\textbf{85.1}&2.9 \\
 & FaSNet-TAC \cite{permutation}              &  2.75M& 2.28 &	74.7 &\textbf{10.0}& 2.40 & 76.7 &\textbf{10.7}& 2.39 & 81.5 &\textbf{10.9}&	2.38  &  76.3 &\textbf{10.6}&	2.40  &	81.5 &\textbf{10.9}\\
     
     & CP-NBDF (prop.)                       & 1.22M & 2.45 &	73.4 &9.3& 2.67  & 78.4 &10.3& 2.61 & 82.5 &10.3&	2.65  &77.9 &9.9&	2.61   &	82.4 &10.1   \\
REAL & CC-NBDF (prop.)                      & 1.33M & 2.43 &	73.4 &9.2& 2.65 & 78.0 &10.3&2.58 & 82.0 &10.3& 2.64 &77.3 &9.9 &	2.59  &	81.7 &10.2 \\ 
    & PW-NBDF (prop.)                      & 1.19M & 2.42 &	74.9 &9.0& 2.68 & 79.2 &10.3& 2.62 & 83.4 &10.3&	2.70 &79.1 &10.1&	2.66     &	83.6 &10.4 \\
     & PW-NBDF with MA (prop.)               & 1.19M & \textbf{2.50} & \textbf{75.1} &9.7& \textbf{2.72} & \textbf{79.3} &10.6& \textbf{2.67} & 83.5 &10.7&	\textbf{2.73} & \textbf{79.2} &10.3&	\textbf{2.70}    	& 83.7 & 10.6  \\ 
     \cmidrule[0.25pt](lr){2-3}  \cmidrule[0.25pt](lr){4-6} \cmidrule[0.25pt](lr){7-9} \cmidrule[0.25pt](lr){10-12} \cmidrule[0.25pt](lr){13-15} \cmidrule[0.25pt](lr){16-18}
     & One-array-dedicated           & 1.20M & 2.56 &	76.7 & 10.1 & 2.68  &80.0 & 10.7& 2.89 &86.9&11.1&	2.73 &	80.2 &10.2&	2.83  & 86.1 & 11.0 \\
     & Microphone-number-dedicated   & 1.20M &  2.43 &	74.6 &9.4& 2.70& 79.4 &10.5& 2.65 &83.8 &10.5&	2.69  &	78.7 &10.1&	2.65  &	83.1 &10.2 \\ 
     \bottomrule[1.25pt]
\end{tabular}
\vspace{-0.3cm}
\end{table*}

\section{Experiment and evaluation }
\vspace{-0.1cm}
\subsection{Experimental setup}
\vspace{-0.1cm}
To collect a large amount of data from different microphone arrays, we use virtual microphone arrays with simulated RIRs and simulated multichannel noise in this work. 
We totally plant 116 virtual microphone arrays with different sizes, geometry and numbers of microphones. For training, 114 virtual arrays are used with microphone numbers in the range of [2, 8] in different (linear, circular, circular with center microphone, nonuniform linear, ad-hoc) shapes. For each shape, there are five different diameters in the range of [0.15, 0.5] m. RIRs are simulated with different room setups using the toolkit \cite{RIR-generator} for each of these arrays. Three virtual rooms are created and each virtual array is randomly positioned at 1m height. The azimuth, the source-array distance and reverberation time (RT60) are randomly chosen in ranges of [$0^{\circ},360^{\circ}$), (0.5, 4.5) m and [0.14, 1] s. As for test, the microphone linear array (LA) presented in \cite{SG} and the circular array (CA) provided in the \emph{REVERB challenge} \cite{REVERB} with real-measured RIRs are used. For LA, the spacing of eight microphones is [4,4,4,8,4,4,4] cm, and the RIRs are measured with two source-array distances, 13 azimuths and three RT60 setups (0.16, 0.36, 0.61) s. For CA, eight microphones form a circle with a diameter of 20 cm, and the RIRs are measured with two source-array distances, two azimuths and three RT60 setups (0.25 s, 0.68 s, 0.73 s) s. Besides the real arrays, we also test the rest two virtual arrays, which have exactly the same geometry as the two real arrays, respectively. The RIRs for the virtual microphones are simulated with RT60 randomly chosen from a range of [0.25, 0.75] s. To have a varying number of microphones, 2-, 4-, 6-ch sub-arrays (excluded in training) are extracted from each SIMU and REAL original 8-ch array, which then generates five types of sub-arrays: 2-CH (from both LA and CA) arrays, 4-CH CA, 4-CH LA, 6-CH partial CA and 6-CH LA.

We create 12000, 2000 and 1200 (600 per sub-array) multichannel noisy reverberant speech for training, validation and test,  respectively. Clean utterances from TIMIT \cite{timit} are convolved with multichannel RIRs.  Multichannel spatially diffuse noise is generated by using \cite{noise-generator} with babble or white noise clips and wind noise from \cite{wind-generator}. Multichannel speech (containing only one single source) and diffuse noise are mixed with a signal-to-noise ratio (SNR) randomly chosen from the range of [-5, 10] dB for training while from [-5, 0] dB for test. STFT is performed with 512 samples-long (32 ms) hamming window and a hop size of 256-samples. Adam is used for optimization with a learning rate of 0.001. Magnitude augmentation (MA) in Section 2.4 is applied for PW-NBDF with multiplication factors randomly chosen from [0.75, 1.33]. Three objective evaluation metrics are adopted: i) perceptual evaluation speech quality (PESQ) \cite{PESQ}; ii) short-time objective intelligibility (STOI) \cite{STOI}; iii) signal-to-distortion ratio (SDR) \cite{SDR}.  

Three baseline methods are compared. FaSNet with TAC\footnote{https://github.com/yluo42/TAC} \cite{permutation} is trained with the same data as for the proposed network. The scale invariant-SNR (SI-SNR) is taken as the training loss. BeamformIt \cite{beamformit} and Oracle MVDR are compared as well where the Oracle MVDR beamformer estimates the steering vector and noise covariance matrix using the clean (reverberant) speech and pure noise of the test signal, respectively.

\vspace{-0.1cm}
\subsection{Results and discussion}
\vspace{-0.1cm}
First, we evaluate the performance while taking the training scheme presented in Section 2.2 for a constant array (the simulated 6-CH LA for test in this experiment). Two setups are tested as benchmarks: \romannumeral1) Six NBDF models trained with fixed permutation by alternately setting different channels as the reference. The evaluation metrics of six models are averaged. This setup, referred to as constant-reference-permutation, uses the same reference channel and channel permutation for training and test, which is supposed to achieve the best performance in the same framework but lack inner-array generalization. 
\romannumeral2) The randomly selected reference channel placed at the first place while the rest are in a constant permutation. The constant-permutation setup allows to alter the reference channel during test, but other microphones follow the same permutation with training. We remind that the proposed training scheme allows altering both the reference channel and the permutation during test. In Table 1, the performance slightly decreases from constant-reference-permutation benchmark to constant-permutation benchmark, then further to the proposed scheme, which demonstrates that the proposed scheme is able to largely increase the flexibility of multichannel speech enhancement model on a fixed array with a tiny loss of performance.

To evaluate the cross array generalization capability, two other benchmark models are tested: \uppercase\expandafter{\romannumeral1}) the basic NBDF network is independently trained for each test microphone sub-array using the proposed training scheme, namely one-array-dedicated. \uppercase\expandafter{\romannumeral2}) Different basic NBDF networks are trained using arrays for each group of arrays with the same number of microphones in training sets, three networks for 2-CH, 4-CH and 6-CH groups respectively. This is referred to as microphone-number-dedicated. This setup allows to work on an unseen array with the same microphone number as the corresponding network. The results are shown in Table 2. Our proposed methods have comparable results on SIMU data, in the sense that their performances are close to, even exceed the one-array-dedicated model especially for PW-NBDF. On REAL data, similar behaviors to SIMU data are observed, which demonstrates that the model trained with virtual microphone arrays and simulated RIRs can generalize to real microphone arrays. Furthermore, by integrating MA, PW-NBDF is consistently improved, which shows the effectiveness of this data augmentation technique. The better performance of PW-NBDF relative to CP-NBDF, probably reveals that global spatial correlation can be somehow represented by dual-channel correlation in the NBDF framework. To achieve good performance however, the inter channel correlation has to be effectively learned, as PW-NBDF with shared LSTM outperforms CC-NBDF with shared CNNs.  

Better performance scores are achieved by both FaSNet-TAC and all of the proposed networks compared to BeamformIt and Oracle MVDR, as shown in Table 2. This demonstrates that learning spatial information with DNNs compared to hand-crafted beamforming techniques is indeed more efficient, for noise reduction,  especially in reverberant and low-SNR environments. Compared to FaSNet-TAC, our proposed \emph{PW-NBDF with MA} with less than half of the model size, and achieves notably better PESQ and STOI scores and comparable SDR scores (despite the FaSNet-TAC's training loss, i.e. SI-SNR,  is extremely similar to SDR). This performance superiority is possibly due to the fact that the proposed network simultaneously learns the temporal-spatial information, which is important for modeling the directional continuity of speakers and the spatial diffuseness of noise. 
\vspace{-0.3cm}
\section{Conclusion}
This paper systematically studies the microphone array generalization problem with respect to reference channel, microphone permutation, array geometry, and magnitude variation between microphones. We have proposed a network training scheme, three narrowband network architectures, and a magnitude augmentation technique, which overall achieves a simple yet effective multichannel speech enhancement network with excellent microphone array generalization performance. Moreover, the proposed network achieves better speech enhancement performance than the oracle beamforming technique and the advanced FaSNet-TAC method. One shortcoming of this work is that the simulated multichannel noise used in the experiments is ideally diffuse. However, the real-recorded multichannel noise has much more complex spatial characteristics, it could be an arbitrary combination of partially diffuse, directional, and spatially-uncorrelated noise. This issue will be studied in future work.

\bibliographystyle{IEEEtran}

\bibliography{mybib}


\end{document}